\documentclass[12pt]{article}
\usepackage{epsfig}
\begin{document}
\begin{titlepage}
\begin{center}

\vspace{2cm}

{\Large \bf Soft Contribution to Quark-Quark Scattering
Induced by an Anomalous Chromomagnetic Interaction}
\vspace{0.50cm}\\
Nikolai Kochelev$^{a,b,}$\footnote{kochelev@theor.jinr.ru}\\
{(a) \it School of Physics and Center for Theoretical Physics,
Seoul National University,
Seoul 151-747, Korea}\\
\vskip 1ex {(b) \it Bogoliubov Laboratory of Theoretical Physics,
Joint Institute for Nuclear Research, Dubna, Moscow region, 141980
Russia} \vskip 1ex
\end{center}
\vskip 0.5cm \centerline{\bf Abstract} We calculate the soft
contribution to high energy quark-quark scattering
that arises from an instanton-induced quark anomalous
chromomagnetic moment. We demonstrate that this is a large
contribution, which cannot be neglected for transverse
momenta of a few GeV. We discuss the influence
of this effect on inclusive particle production.
\vspace{1cm}
\end{titlepage}
\setcounter{footnote}{0}

\section{Introduction}

The theoretical interpretation of the precise data now available
on inclusive particle production is one of the central problems
of QCD. These data provide important information on the
nature of the strong interaction between quarks and gluons
for a range of scattering kinematics, and their
interpretation is very useful for a more complete
understanding of QCD.
Furthermore, inclusive particle production data is used
as a baseline for the interpretation of more complicated
experimental data, such as secondary particle correlations,
spin asymmetries, and so forth. One specific area of particular
interest is the possibility of studying the properties of
quark-gluon matter produced in heavy ion collisions at RHIC,
through the use of inclusive particle production rates.

One of the main components of perturbative
QCD (pQCD) calculations of inclusive rates is the set of
perturbative partonic cross sections. However, it has been found
that the use of the leading-twist partonic cross sections in successive
orders of the strong coupling constant $\alpha_s$ (LO, NLO, etc.)
does not lead to a satisfactory description of the available
high energy inclusive data over the broad kinematical range
where pQCD is expected to be applicable \cite{aurenche,soffer,brodsky}.
One possible resolution of this discrepancy is to
include the effects of intrinsic parton transverse momentum
distributions. (For a recent discussion see Ref.\cite{murgia}.)
This approach allows one to incorporate higher-twist corrections
to pQCD predictions for the inclusive rates in the context of a specific
phenomenological model of transverse momenta.

Unfortunately, the non-zero intrinsic transverse momenta of the
initial partons implies the loss of the advantages of collinear
pQCD calculations. Furthermore, it is difficult to justify
the large value of the average intrinsic transverse
momentum of partons in the nucleon which is required to fit the
inclusive data,
$<k_\bot>\approx 1$ GeV.
This large value disagrees with the scale of
$<k_\bot>\approx 200$ MeV $\approx 1/R_{conf}$ one would expect
if the origin of the intrinsic momentum was simply confinement.

Here we consider another important higher-twist
contribution to partonic cross sections, which is related to the
topological structure of the QCD vacuum.
It is well known that the QCD vacuum has a rich spectrum of
gluon field fluctuations, with different length and momentum scales.
One type of long-range gluonic excitations, for example, is
responsible for confinement. Another important category includes
the strong topological fluctuations known as instantons;
these play significant roles in hadron physics, such as providing
a mechanism for chiral and $U(1)_A$ symmetry violations
\cite{shuryak,diakonov}.

Instantons have a small typical size of $\rho_c\approx 0.3$ fm,
and can therefore provide higher-twist contributions which remain
significant at the momentum transfers considered here. In particular,
instantons lead to a nonperturbative anomalous chromomagnetic
quark-gluon vertex, which is not encountered in conventional pQCD
\cite{diakonov,kochelev}.
In this letter we demonstrate that the contribution of this
quark anomalous chromomagnetic moment to high energy quark-quark
scattering is large in this kinematic regime, and is
therefore an interesting candidate for the fundamental QCD mechanism
underlying the important higher-twist effects observed in inclusive
particle production.

\section{Chromomagnetic contribution to quark-quark scattering}

The effective chromomagnetic quark-gluon interaction induced by
instantons is given by \cite{diakonov,kochelev}
\begin{equation}
{\cal L}_{eff}= -i\frac{g_s\mu_a}{2M_q} \bar q
\sigma_{\mu\nu}t^aG^a_{\mu\nu}q,
\label{chrom}
\end{equation}
where $\mu_a$ is the quark anomalous
chromomagnetic moment,  $M_q$ is the effective quark mass in the
instanton vacuum, and $G^a_{\mu\nu}$ is the gluon field strength tensor.
For an off-shell gluon of virtuality $q$, the instanton-induced
quark-gluon vertex of (\ref{chrom}) should be multiplied by the
instanton form factor
\begin{equation}
F(z)=\frac{4}{z^2}-2K_2(z),
\label{form2}
\end{equation}
where $z=q\rho_c$. The value of the quark anomalous chromomagnetic
moment is proportional to the instanton packing fraction in the
QCD vacuum, $f=n_c\pi^2\rho_c^4\approx 0.1$, where $n_c$ is the
instanton density.

The diagram which gives the instanton contribution
to quark-quark scattering in the high energy limit
($s >> |t|$) due to the interaction (\ref{chrom}),
at leading order in the instanton packing fraction,
is shown in Fig.1.
(Here, $s=(p_1+p_2)^2$ and $t=(p_1-p_3)^2$.)
A straightforward calculation gives the resulting
differential cross section,
\begin{equation}
\frac{d\sigma^{chrom}}{dt} =
\frac{2\mu_a^2|t|(F(\sqrt{|t|}\rho_c))^2}{M_q^2}
\frac{d\sigma^{pert}}{dt} \label{cs}
\end{equation}
where
\begin{equation}
\frac{d\sigma^{pert}}{dt} =
\frac{8\pi\alpha_s(t)}{9t^2}
\label{pqcd}
\end{equation}
is the LO one-gluon pQCD cross section.
\begin{figure}[htb]
\begin{minipage}[c]{8cm}
\vspace*{-0.0cm}
\centerline{\epsfig{file=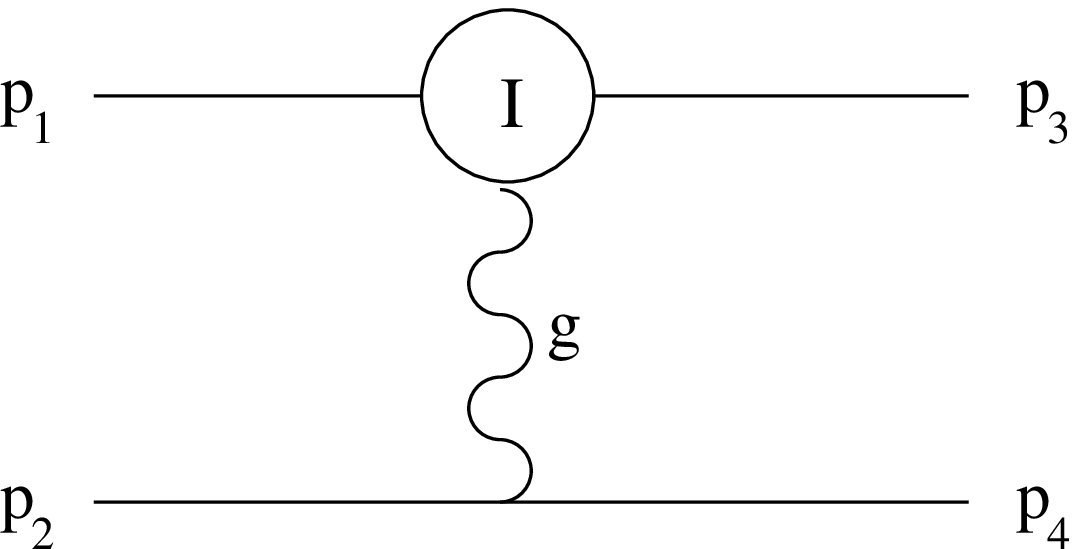,width=7cm,angle=0}}\ \caption{
The Feynman diagram representing the contribution of the
quark chromomagnetic moment to high energy
quark-quark scattering.}
\end{minipage}
\hspace*{0.5cm} \vspace*{0.5cm}
\begin{minipage}[c]{8cm}
\centering \centerline{\epsfig{file=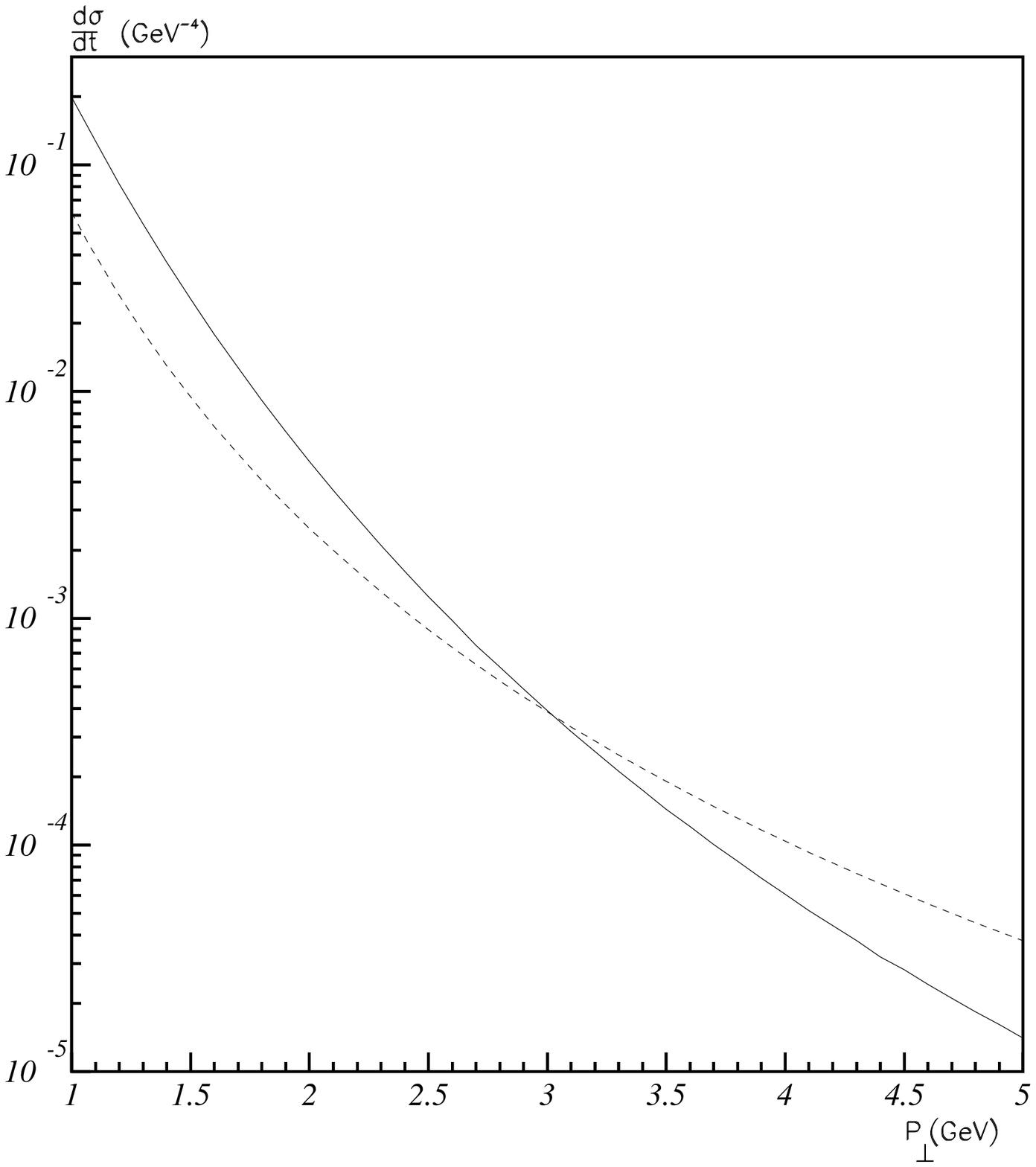,width=8cm,angle=0}}\
\caption{Perturbative (dashed) and nonperturbative (solid)
quark-quark differential cross sections versus transverse momentum.}
\end{minipage}
\end{figure}

In Fig.2 we compare the pQCD (one-gluon exchange) and
nonperturbative chromomagnetic results for the
elastic quark-quark scattering cross section
as a function of transverse momentum, for the special case of
$90^\circ $ scattering ($p_\bot=\sqrt{-t/2}$).
We assume a value of
\begin{equation}
\mu_a=-0.744,
\end{equation}
for the quark anomalous chromomagnetic moment, which
was obtained in Ref.\cite{diakonov} for the case
$N_f=3$.~\footnote{In the calculation of the instanton-induced
quark anomalous chromomagnetic moment reported in Ref.\cite{kochelev}
for the case $N_f=1$, a much smaller value of $\mu_a\approx -0.2$ was found.}
We also assume a constituent quark mass of
$M_q=350$ MeV, and an instanton inverse length scale of
${\rho_c}^{-1}=600$ MeV;
these are typical of the values used in the instanton liquid model
\cite{shuryak,diakonov}.
For the strong coupling constant, the following parametrization
was used for the case $N_f=3$;
\begin{equation}
\alpha_s(t)=\frac{4\pi}{9\ln((|t|+4m_g^2)/\Lambda^2)},
\end{equation}
where $m_g^2=0.2$ GeV$^2$ and $\Lambda=0.16$ GeV \cite{brodsky2}.
One can see in Fig.2 that this nonperturbative contribution is
quite large, and exceeds the pQCD result for transverse
momenta of $p_\bot < 3$ GeV. This suggests that the pQCD regime
is only encountered at rather larger values of $p_\bot$.

The asymptotic behaviour of the nonperturbative contribution for
large $p_\bot$ is determined by the instanton form factor (\ref{form2}),
\begin{equation}
\frac{d\sigma^{chrom}}{dp_\bot}\approx \frac{A}{p_\bot^6},
\label{pt}
\end{equation}
where $A$ is a constant. This $p_\bot$ dependence is steeper than is
predicted by leading-twist pQCD, which anticipates
$1/p_\bot^4$. We note in passing that
the value $n^{chrom}=6$ in (\ref{pt}) is in agreement with
the result $n_{eff}=6.33\pm 0.54 $ found in the
recent analysis of RHIC data
on inclusive neutral pion
production in peripheral heavy ion collisions \cite{RHIC}.
(See also the discussion in Ref.\cite{brodsky}.)
Thus it appears that this nonperturbative scattering mechanism
may also provide a viable explanation for
the recent RHIC inclusive pion production data.

\section{Conclusion}
We have discussed a novel nonperturbative, instanton-based
contribution to quark-quark scattering at high energies. This
scattering mechanism is due to the existence of a quark anomalous
chromomagnetic moment, which originates from the instanton
structure of the QCD vacuum. We have shown that this new soft
contribution to the partonic cross section is numerically rather
large in the kinematic regime considered, and may be responsible
for the discrepancy between leading-twist pQCD predictions and the
existing high energy data on inclusive particle production. We
should also emphasize  that due to its strong helicity dependence,
a large contribution  coming from an anomalous chromomagnetic
interaction to a single spin \cite{kochelev,hoyer} as well to a
double spin asymmetries is expected.

\section{Acknowledgments}
We are happy to acknowledge useful discussions of various aspects
of this research with A.E.Dorokhov, and we are especially grateful
to Prof. D.-P. Min for his kind hospitality at the School of
Physics of Seoul National University in the final stages of this
work. We would like to thank T.Barnes for careful reading the
manuscript. This work was supported in part by the Brain Pool
program of the Korea Research Foundation through KOFST grant
042T-1-1, and by the Russian Foundation for Basic Research through
grant RFBR-04-02-16445.

\end{document}